\documentclass[debug]{rmaa}
\usepackage{natbib}
\usepackage[utf8]{inputenc}

\usepackage{paralist}

\usepackage{psfrag,color}
\usepackage{url}
\usepackage{xcolor}
\definecolor{newcolor}{rgb}{.8,.349,.1}
\newcommand{\source}{XMMSL1 J171900.4--353217}
\newcommand{\swift}{\textit{Swift}}
\newcommand{\nicer}{\textit{NICER}}
\newcommand{\xmm}{\textit{XMM-Newton}}
\newcommand{\integral}{\textit{INTEGRAL}}
\newcommand{\rxte}{\textit{RXTE}}

\title{X-ray observations of the very-faint X-ray transient XMMSL1 J171900.4--353217: A new candidate neutron star low-mass X-ray binary} 

\shorttitle{AHMED, DEGENAAR, WIJNANDS \& ARMAS PADILLA}

\author{
  O. Ahmed,\altaffilmark{1,2} 
  N. Degenaar,\altaffilmark{3}
  R. Wijnands,\altaffilmark{3}
  and M. Armas Padilla\altaffilmark{4,5}}

\altaffiltext{1}{Astronomy and Space Science Department, Faculty of Science, King Abdulaziz University, P.O. Box 80203, Jeddah 21589, Kingdom of Saudi Arabia}

\altaffiltext{2}{Department of Physics, Faculty of Natural and computational Sciences, Debre Tabor University, P.O. Box 272, South Gondar, Ethiopia}

\altaffiltext{3}{Anton Pannekoek Institute for Astronomy, University of Amsterdam, Science Park 904, 1098 XH, Amsterdam, The Netherlands}
\altaffiltext{4}{Instituto de AstrofÃsica de Canarias, 38205, San Cristobal de La Laguna, Spain}
\altaffiltext{5}{Departamento de Astrofísica, Universidad de La Laguna, E-38206 La Laguna, Tenerife, Spain}
\shortauthor{Ahmed, Degenaar, Wijnands \& Armas Padilla}
\fulladdresses{
\item Nathalie Degenaar and Rudy Wijnands: Anton Pannekoek Institute for Astronomy, University of Amsterdam, Science Park 904, 1098 XH, Amsterdam, The Netherlands.
\item Osman Ahmed: Astronomy and Space Science Department, Faculty of Science, King Abdulaziz University, P.O. Box 80203, Jeddah 21589, Kingdom of Saudi Arabia\\
Department of Physics, Faculty of Natural and computational Sciences, Debre Tabor University, P.O. Box 272, South Gondar, Ethiopia
\item Montserrat Armas Padilla: \item Instituto de AstrofÃsica de Canarias, 38205, San Cristobal de La Laguna, Spain\\
Departamento de Astrofísica, Universidad de La Laguna, E-38206 La Laguna, Tenerife, Spain
}

\listofauthors{Ahmed, Degenaar, Wijnands and Armas Padilla}
\indexauthor{Ahmed, O.}
\indexauthor{Nathalie, D.}
\indexauthor{Wijnands, R.}
\indexauthor{Armas Padilla, M.}

\abstract{\source\ is a very-faint X-ray transient that was discovered in 2010 March when it exhibited an outburst. We report on 7 observations, obtained with the X-Ray Telescope (XRT) aboard the Neil Gehrels {\it{Swift}} Observatory between 2010 May to October. %We study the spectral evolution of the outburst with the aim of gaining insight into the nature of the central accretor.
By fitting a single absorbed power-law model to the XRT spectra, we infer power-law indices of $\Gamma = 1.8-2.7$ and an absorption column density of $\mathrm{N}_{\mathrm{H}}=(4.6-7.9)\times 10^{22}~\mathrm{cm}^{-2}$. 
The inferred $0.5-10$~keV luminosity fluctuated irregularly and peaked at $\mathrm{L}_\mathrm{X}\simeq 10^{35}-10^{36} ~\mathrm{erg~s^{-1}}$ for a distance of $4-12$~kpc. Based on the evolution of the power-law index with varying luminosity, we propose that the source most likely is a transient neutron star low-mass X-ray binary located at several kpc. If true, it would be a good candidate to search for coherent millisecond pulsations when it enters a new accretion outburst. }

 \resumen{XMMSL1 J171900.4-353217 es una binaria de rayos-X transitoria poco luminosa descubierta en marzo de 2010 durante una erupci\'{o}n. Presentamos 7 observaciones obtenidas entre mayo y octubre de 2010 con el Telescopio de Rayos-X (XRT) a bordo del Observatorio Neil Gehrels Swift. Mediante el ajuste de los espectros del XRT con un modelo de ley de potencias absorbido, obtenemos un \'{i}ndice fot\'{o}nico de $\Gamma = 1.8-2.7$ y una densidad de la columna de hidr\'{o}geno de $\mathrm{N}_\mathrm{H}=(4.6-7.9)\times 10^{22}$ \, ${\mathrm{cm}}^{-2}$. La luminosidad, en el rango $0.5-10$ \, keV, fluctu\'{o} irregularmente, con picos de L$_\mathrm{X} \approx 10^{35} - 10^{36}$ \, {erg~ s}$^{-1}$ para una distancia de $4-12$ \, kpc. Bas\'{a}ndonos en la evoluci\'{o}n del \'{i}ndice fot\'{o}nico con la luminosidad, proponemos que la fuente es probablemente una binaria de rayos-X poco masiva con estrella de neutrones situada a varios kpc.
 De ser cierto, esta fuente ser\'{i}a una buena candidata para buscar pulsaciones coherentes de milisegundos cuando entre de nuevo en erupci\'{o}n.
}

\addkeyword{ISM: abundances}
\addkeyword{(stars:) binaries: general}
\addkeyword{stars: individual (\source\ )}
\addkeyword{stars: neutron}
\addkeyword{X-rays}

\begin{document}
% % \day{Received November 14 2023; accepted August 15 2024}
% % Typeset article header
% \newcommand{\custommaketitle}{
%     \begin{center}
%         \textit{Revista Mexicana de Astronomía y Astrofísica, 60, 403–412 (2024)\\
%         c© 2024: Instituto de Astronomía, Universidad Nacional Autónoma de México\\
%         https://doi.org/10.22201/ia.01851101p.2024.60.02.18}
%     \end{center}
% \vspace{1cm}
\maketitle
%%%
% \begin{document}
% \custommaketitle

\section{Introduction}\label{sec:intro}

X-ray binaries are binary systems in
which a compact object, either a black hole (BH) or neutron star (NS), accretes matter from a companion star. When the companion is a low-mass star ($\mathrm{M}\lesssim 1 \mathrm{M}_{\odot}$), the system is known as a low-mass X-ray binary (LMXB). Many LMXBs are transient: they become bright only during outbursts of active accretion, but are more often found in a dim quiescent state.

% THE TEXT BELOW DOES NOT FIT HERE -- MOVED TO THE DISCUSSION
%For stellar mass companions to achieve very low masses at the beginning, standard LMXBs evolution is very slow to achieve that very faint X-ray transients probably either have hydrogen poor companions or born with low companion masses \citep[][]{King2006}, and  During their study of X-ray transients near the Galactic center \citep[][]{degenaar2009behavior} for accreting NS and BH X-ray binaries the time-average accretion rate, $<\dot{M}_{\mathrm{long}}>$, of the transients was a significant component in explaining the nature of very faint X-ray transients.

% These are often transient, meaning that they are most of their time in a quiescent state during which no or hardly any accretion occurs and consequently their X-ray (2--10 keV) luminosities are low, typically $\simeq 10^{30}-10^{33} ~{\mathrm{erg/s}}$ \citep{Wijnands2015}. 
In quiescence, LMXBs have a low X-ray luminosity of $\mathrm{L}_\mathrm{X}\lesssim^{33}~\mathrm{erg~s^{-1}}$ \citep[e.g.,][]{wijnands2017}.
The maximum luminosity that is reached in outburst can vary a lot from source to source, and even from outburst to outburst for a single object. While many LMXBs are bright, with 2--10 keV peak luminosities of $\mathrm{L}_\mathrm{X}\simeq10^{37}-10^{39}~\mathrm{erg~s^{-1}}$, some also exhibit 'mini outbursts' that reach much lower peak luminosities of $\mathrm{L}_\mathrm{X}\simeq10^{34}-10^{36}~ \mathrm{erg~s^{-1}}$ \citep[e.g.,][]{degenaar2009behavior,wijnands2013sax,cotizelati2014,zhang2019}. These are often shorter than regular bright outbursts, although there are also LMXBs that accrete at such a low-luminosity level for extended periods of time \citep[][]{simon2004,Degenaar2014,allen2015,parikh2018}.

Interestingly, a growing number of systems has been discovered that exhibit maximum outburst luminosities of $\mathrm{L}_\mathrm{X}\simeq10^{34}-10^{36}~ \mathrm{erg~s^{-1}}$ and seemingly never exhibit brighter outbursts \citep{sakano2005,muno2005-vfxts,degenaar2009behavior,bozzo2015,bahramian2021}. These LMXBs belong to the class of very faint
X-ray transients \citep[VFXTs;][]{Wijnands2006}. Many of these VFXTs are found near the Galactic center, but this is very likely a selection bias since this region has been regularly surveyed by many X-ray missions hence the brief and dim outbursts of VFXTs are more easily discovered than in other regions of our Galaxy \citep[e.g.,][]{muno2005-vfxts,sakano2005,Wijnands2006,degenaar2015}.

While VFXTs could be intrinsically brighter sources located at large distances (tens of kpc) within Milky Way, estimates from thermonuclear bursts\footnote{Thermonuclear burst, or type-I bursts, are brief (seconds to hours) flashes of X-ray emission caused by unstable nuclear burning of gas accreted onto a neutron star. These explosions are thought to reach the Eddington limit and can therefore be employed to derive a distance to the bursting LMXB \citep[e.g.,][]{kuulkers2003}. exhibited by many VFXTs place them at distances of only several kpc and they must thus have low intrinsic luminosities \citep[e.g.,][]{cornelisse2002,lutovinov2005,degenaar2010_burst,bozzo2015,keek2017}}. In addition, while inclination effects could possibly make these systems appear fainter than they intrinsically are \citep[e.g.,][]{Muno2005}, this can likely only account for a small fraction of the VFXTs \citep[see][]{King2006}. 
Many VFXTs are thus expected to be intrinsically faint, i.e. accrete at low rates. This makes them interesting for a number of scientific reasons. For instance, they probe a little explored mass-accretion regime, hence are valuable for studying accretion physics
\citep[e.g.,][]{2013MNRAS.434.1586A,2015MNRAS.447..486W,2017MNRAS.464..398D}. In addition, VFXTs are interesting for testing and improving binary evolution models \citep[e.g.][]{King2006, degenaar2010four,2015arXiv150102769M}, and for increasing our understanding of nuclear burning on the surface of
accreting NS \citep[e.g.,][]{2007ApJ...654.1022P,2010MNRAS.404.1591D}.

Despite that the number of VFXTs has now grown to a few tens of systems \citep[e.g.,][]{bahramian2023} and detailed studies of several systems have been performed over the past decade, still much remains to be learned about this source class. For instance, there is no clear picture yet about the distribution of system properties such as the nature of the compact accretor, type of companion star, and the size of the orbit. Determining whether an LMXB harbors a NS or a BH requires direct measurements of the physical properties of the compact object, such as its mass, or to detect the presence of a solid surface (e.g., through X-ray pulsations or thermonuclear bursts). However, such measurements are often challenging for VFXTs due to their faintness \citep[e.g., making pulsation searches challenging;][]{vdeijnden2018} and low accretion rates \citep[e.g., rendering thermonuclear bursts rare;][]{degenaar2011}. 

For some VFXTs, indirect approaches of using the ratio between the X-ray and radio or optical/infrared luminosity have been employed to assess the nature of the compact accretor \citep[e.g.,][]{armaspadilla2011,paizis2011}. However, their short outbursts often make it difficult to identify a counterpart for VFXTs at other wavelengths \citep[e.g.,][]{shaw2020}. Furthermore, due to their low accretion rates, not many VXFTs have been detected in the radio band \citep[][]{vandeneijnden2021} and for finding optical/infrared counterparts added complications arise from their biased locations in the direction of the Galactic center \citep[i.e., high extinction and crowding; e.g.,][]{Bandyopadhyay2005}. Fortunately, an indication of the nature of the accretor can also be obtained by studying the X-ray spectral evolution of VFXTs \citep[][]{2011MNRAS.417..659A,beri2019,stoop2021}.

The X-ray properties of LMXBs harboring a NS can be very similar to those containing a BH. However, when comparing their X-ray spectra at low luminosities of $\mathrm{L}_\mathrm{X} \simeq10^{34}-10^{36} ~{\mathrm{erg~s^{-1}}}$, it turns out that confirmed or candidate BH systems have significantly harder spectra than confirmed NSs. In addition, the BH spectra show a strong softening when the X-ray lumninosity evolves below $\simeq 10^{34} ~{\mathrm{erg~s^{-1}}}$, while NSs start to show clear softening already at higher X-ray luminosities of $\mathrm{L}_\mathrm{X} \simeq 10^{36} ~{\mathrm{erg~s^{-1}}}$ \citep[e.g.,][]{Wijnands2015, 2017MNRAS.468.3979P}.

Over the last few years, detailed studies have been performed for a growing number of VFXTs and the general conclusion is that due to low statistics on their X-ray spectra, such systems can be satisfactorily described with a simple power-law model, with a soft (black body) component only being distinguishable when high quality (i.e. many counts) data are available \citep[e.g.,][]{2011MNRAS.417..659A}. However, irrespectively of what model is fitted to the spectra, VFXTs also
become softer with decreasing X-ray luminosity. 
%This appears to be true for both BH and NS systems \cite[see e.g.,][]{2011MNRAS.417..659A, 2013MNRAS.434.1586A}, however When the X-ray luminosity evolves below $10^{34} ~{\mathrm{erg/s}}$, BH spectra exhibit strong softening, whereas NSs begin to exhibit softening at higher X-ray luminosities of $10^{36} ~{\mathrm{erg/s}}$ \citep[see][]{Wijnands2015, 2017MNRAS.468.3979P}. 
Their X-ray spectral evolution during an outburst can thus be used as a diagnostic for the nature of the compact accretor.

\subsection{Discovery of \source}\label{subsec:source}
\source\ was discovered as an X-ray transient in \textit{XMM-Newton} slew data obtained on 2010 March 10 %, with a 0.2--10 keV count rate of 4.5 $\mathrm{ct/s}$ (EPIC-pn detector;)
\citep[][]{2010ATel.2607....1R}. The source location was in FOV of \textit{INTEGRAL} observations performed around the same time, but it was not detected %in the \textit{IBIS/ISGRI} mosaic 
\citep[20--40 keV;][]{2010ATel.2616....1B}. \citet{2010ATel.2615....1M} pointed out that \source\ was likely associated to a faint transient source, XTE J1719--356,
%, or the two transients are simultaneously active, 
detected in \textit{RXTE}/PCA scans of the Galactic bulge since 2010 March. 

Observations performed with the X-Ray Telescope \citep[XRT;][]{2005SSRv..120..165B} onboard the \textit{Neil Gehrels Swift Observatory} \citep[\swift;][]{2004ApJ...611.1005G} in 2010 May, showed that the source was still active, i.e. two months after the initial discovery \citep[][]{2010ATel.2627....1R}. \citet{2010ATel.2656....1A,2010ATel.2738....1A} reported on further \textit{Swift}/XRT observations, performed in 2010 May and June, which showed that the source remained active in soft X-rays albeit with varying flux. While \textit{Swift}/XRT did no longer detect the source in 2010 July, suggesting it had returned to quiescence \citep[][]{2010ATel.2722....1A}, \textit{INTEGRAL} serendipitously detected the source in hard X-rays in August 2010 \citep[20--40 keV;][]{2010ATel.2803....1I}. It was also detected again in soft X-rays with \textit{Swift}/XRT around that time \citep[][]{2010ATel.2807....1P}. Nothing more was reported on the source after this.

In this work we investigate the nature of the compact accretor in the VFXT \source\ by studying its X-ray spectral evolution as seen with \textit{Swift}/XRT.

\section{Observations and data analysis}\label{sec:obs}

\source\ was observed over a 157 days time-span with {\it Swift}, between 2010 May 11 and October 15 (see Table \ref{Log of}). Seven pointed observations were carried out during this time and we investigate the data collected using the XRT.

\begin{table}[!t]\centering
  \setlength{\tabnotewidth}{0.5\columnwidth}
  \tablecols{3}
  % Stretch the space between table columns 
  \setlength{\tabcolsep}{0.9\tabcolsep}
  \caption{Log of \textit{Swift}/XRT observations.}\label{Log of}
  \begin{tabular}{lcccc}
 % \begin{tabular}{l @{\DS} ccccc}
    \toprule
Obs & Observation ID & Date and start time & Exposure time & Net count rate \\
&& (UT)&  (ks) & (${\mathrm{ct}~\mathrm{s}^{-1}}$)
 \\
\midrule
            1&00031719001 & 2010 May 11 16:56&   2.0 &0.4 \\ 
			2&00031719002 & 2010 May 31 12:23&    1.0 &0.1 \\ 
			3&00031719003 & 2010 June 14 11:44&   1.2 & 0.2\\ 
			4&00031719004 & 2010 June 29  06:42&   1.7 &0.7 \\ 
			5&00031719005 & 2010 July 13 09:29&   1.0 &$<$0.008 \\ 
			6&00031719006 & 2010 August 20 13:00&  2.5 & 0.3 \\ 
			7&00031719007 & 2010 October 15 02:10&   1.3 &$<$0.002 \\
\bottomrule
%\tablenotetext{a}{Background corrected count rates}
\end{tabular}

\end{table}

\begin{table*}[!t]
\centering
  \small
  \newcommand{\DS}{\hspace{6\tabcolsep}} %% Expanded Space between
  %% some cols\
  
  \begin{changemargin}{-2cm}{-2cm}
    \caption{Results from analysing the \textit{Swift}/XRT spectra.} \label{tab:spec}
    \setlength{\tabnotewidth}{0.95\linewidth}
    \setlength{\tabcolsep}{0.7\tabcolsep} \tablecols{10}
    \begin{tabular}{l @{\DS} cccc l ccc}
      \toprule
Obs & $\mathrm{N}_{\mathrm{H}}$& $\Gamma$&  $\mathrm{F}_{\mathrm{X,abs}}$ & $\mathrm{F}_{\mathrm{X,unabs}}$ & $\mathrm{L}_{\mathrm{X}}$ 4 kpc & $\mathrm{L}_{\mathrm{X}}$ 8 kpc& $\mathrm{L}_{\mathrm{X}}$ 12 kpc\\
& ($10^{22}{\mathrm{cm}}^{-2}$)&& \multicolumn{2}{c}{($10^{-11}~ {\mathrm{erg~cm}}^{-2}{\mathrm{s^{-1}}}$}) &  \multicolumn{3}{c}{($10^{35} ~{\mathrm{erg~s^{-1}}}$)}\\
\midrule
1 & $5.45^{+0.64}_{-0.61}$ & $1.90\pm0.19$ & $4.17\pm0.19$ & $9.33^{+1.63}_{-1.20}$  & $1.78^{+0.32}_{-0.23}$&$7.14^{+1.25}_{-0.92}$ & $16.07^{+2.81}_{-5.07}$\\ 
2 & $5.81^{+2.35}_{-1.98}$ & $2.49^{+0.78}_{-0.70}$ & $0.57^{+0.12}_{-0.09}$ & $2.29^{+4.02}_{-1.03}$ & $0.44^{+0.76}_{-0.20}$& $1.73^{+3.08}_{-0.79}$&$3.94^{+6.93}_{-1.77}$\\ 
3 &$7.85^{+1.87}_{-1.77}$ & $2.73\pm0.52$ & $2.04^{+0.25}_{-0.22}$ & $12.02^{+13.09}_{-5.26}$ & $2.30^{+2.50}_{-1.01}$& $9.20^{+10.02}_{-4.02}$& $20.7^{+22.55}_{-9.06}$\\ 
4 & $4.60^{+0.47}_{-0.45}$ & $1.77\pm0.15$ & $7.01^{+0.40}_{-0.25}$ & $13.80^{+1.69}_{-1.21}$ &$2.64^{+0.32}_{-0.23}$&$10.75^{+1.29}_{-0.93}$ & $23.77^{+2.91}_{-2.09}$\\ 
5 & 5.81 fix & 2.49 fix & $<0.06$ & $<0.24$  &$<0.05$ & $<0.19$ & $<0.42$\\ 
6 & $6.67^{+0.89}_{-0.86}$ & $1.80\pm0.22$ &$2.95\pm0.20$ & $6.61^{+1.33}_{-0.86}$ & $1.26^{+0.26}_{-0.16}$&$5.06^{+1.02}_{-0.66}$ & $11.38^{+2.29}_{-1.48}$\\ 
7 & 5.81 fix & 2.49 fix & $<0.07$ & $<0.27$ & $<0.06$ & $<0.21$ & $<0.47$\\  
\bottomrule
\end{tabular}
%\tablenotetext{}{Quoted errors reflect 1-$\sigma$ confidence intervals.}
% \tabnotetext{}{\small Note: X-ray fluxes and luminosities are given in the 0.5--10 keV energy band and quoted errors reflect 1-$\sigma$ confidence intervals.}
% \tablenotetext{b}{Flux (in units of $10^{-12}~ {\mathrm{erg/cm}}^{2}/{\mathrm{s}}$ over the X-ray energy 0.5 -- 10 keV)}
% \tablenotetext{c}{nhp(null hypothesis probability)-probability that the model is correct for those data points (if close to unity)}
%\tablenotetext{b}{Flux in units of $10^{-11}~ {\mathrm{erg/cm}}^{2}/{\mathrm{s}}$}
%\tablenotetext{c}{X-ray luminosity in units of $10^{35} ~{\mathrm{erg/s}}$ calculated from the unabsorbed flux by adopting distances of 4 kpc, 8 kpc and 12 kpc.}
\end{changemargin}
\end{table*}

\begin{table*}[!t]\centering
  \small
  \newcommand{\DS}{\hspace{6\tabcolsep}} %% Expanded Space between
  %% some cols
  \begin{changemargin}{-2cm}{-2cm}
    \caption{Other reported X-ray flux measurements.}\label{Log}
    \setlength{\tabnotewidth}{0.95\linewidth}
    \setlength{\tabcolsep}{0.6\tabcolsep} \tablecols{6}
    \begin{tabular}{l @{\DS} cccc l c}
      \toprule
Observatory   &Date  & Reported brightness & $\mathrm{F}_{\mathrm{X,abs}}$\tablenotemark{a} &  $\mathrm{F}_{\mathrm{X,unabs}}$\tablenotemark{a}  & Reference\tablenotemark{b} \\ 
(detector)& & (various formats) & \multicolumn{2}{c}{($10^{-11}~ {\mathrm{erg~cm}}^{-2}{\mathrm{s}^{-1}}$)} &  \\
    \midrule
    \textit{XMM-Newton} (PN) &2010 March 10 &  4.5 c$~\mathrm{s}^{-1}$ (0.2--10 keV) & 4.5 & 10.2 & 1\\
%
    %\textit{XMM-Newton}&2010 March 10 & 55266 &7.9& 11.5  &2.2 & 8.8& 19.8\\
    %\textit{INTEGRAL} &2010 March 09 11:17& 55264& 17.5 &34.6 &6.6 & 26.5& 59.6\\
    %\textit{INTEGRAL}  &2010 March 09 11:17&55264 & 7.8 &15.4& 2.9 &11.8& 26.5\\
    \textit{INTEGRAL} (IBIS) & 2010 March 09 & $<$6 mCrab (20--40 keV) & $<6.82$ & $<15.40$ & 2\\
    \textit{INTEGRAL} (IBIS)& 2010 August 14 & $3.0 \times 10^{-11}~\mathrm{erg~cm^{-2}}~\mathrm{s}^{-1}$ (20--40 keV) & 3.55& 7.94 & 3\\ 
 %    && &&&&&&\\
    %\textit{INTEGRAL} &2010 August 14 18:20 &55423 &3.2& 7.3\\ 
   % \textit{INTEGRAL} &2010 August 14 18:20 &55423 &10.7& 21.3& 4.1 & 16.3& 36.7\\
    %\textit{INTEGRAL} &2010 August 20 14:28&55429 &16.9& 33.4 & 6.4& 25.6& 57.5\\
    \textit{INTEGRAL} (IBIS)& 2010 August 20 & $<2.7 \times10^{-11}~\mathrm{erg~cm^{-2}}~\mathrm{s}^{-1}$ (20--40 keV) & $<$3.2 & $<$7.1 & 4\\ 
  %   &&&&&&&&\\
    %\textit{INTEGRAL} &2010 August 20 14:28& 55429 &4.3& 8.4& 1.6& 6.4& 14.5\\
\bottomrule
\tabnotetext{a}{\small The quoted count rates were converted to 0.5--10 keV fluxes using \textsc{webpimms} and assuming a power-law spectral model. For the first two table entries we used $\mathrm{N}_{\mathrm{H}} = 5.45 \times 10^{22}~{\mathrm{cm}}^{-2}$ and $\Gamma = 1.90$, for the last two $\mathrm{N}_{\mathrm{H}} = 6.67 \times 10^{22}~{\mathrm{cm}}^{-2}$ and $\Gamma = 1.80$. These parameter values match those found from our spectral fitting of \swift/XRT data obtained around that time (see Table~2).}
\tabnotetext{b}{References: 1=\citet{2010ATel.2607....1R}, 2=\citet{2010ATel.2616....1B}, 3=\citet{2010ATel.2803....1I}, 4=\citet{2010ATel.2807....1P}. }
%\tablenotetext{b}{Flux in units of $10^{-11}~ {\mathrm{erg/cm}}^{2}/{\mathrm{s}}$}
%\tablenotetext{b}{The source was not detected and an upper-limit flux was calculated from the count rate.}
% \tablenotetext{}{All the data of the table was recorded from \textit{INTEGRAL} an\tablenotetext{a}{Flux in units of $10^{-11}~ {\mathrm{erg/cm}}^{2}/{\mathrm{s}}$}d \textit{XMM-Newton} available with Astronomers Telegram numbers 2803, 2616, 2807 and 2607.}

\end{tabular}
\end{changemargin}
\end{table*}

\subsection{Description of the data reduction}\label{subsec:descri}

All XRT data were collected in photon counting (PC) mode. We reduced the data and
 	obtained science products using the {\sc heasoft} software package
 	(v. 6.26). We cleaned the data by running the {\textsc{xrtpipeline}} task in which standard event grades of 0--12 were selected.
  For every observation, images, count rates and spectra were obtained with the {\textsc{xselect}} (v.2.4) package. 
 We extracted the source events using a circular region with a radius of 52 arcseconds. The background emission was averaged over three circular regions of similar size that were placed on nearby, source-free parts of the image.

The source was detected in 5 of the 7 observations (see Table \ref{Log of}) and for these we extracted spectra. Using {\textsc{grppha}}, three spectra were grouped to have 10 counts per energy bin, one was grouped to 20 counts per bin (observation ID 00031719004, when the source was brightest) and one to 5 counts per bin (observation ID 00031719002 when the source was faintest).
 
  The spectra were corrected for the fractional exposure loss due
 to bad columns on the CCD. For this, we created exposure maps
 with the {\textsc{xrtexpomap}} task, which were then used as an input to generate
 the ancillary
 response files (arf) with the {\textsc{xrtmkarf}} task. We acquired the response matrix file (rmf) from the {\textsc{heasarc}} calibration database (v.12).

\subsection{Pile-Up}\label{subsec:pile}

Observation 00031719004 has the highest count rate
 (0.7 $\mathrm{ct~s^{-1}}$) and is affected by pile-up. We tested this following the steps outlined in the dedicated XRT analysis thread\footnote{\url{https://www.swift.ac.uk/analysis/xrt/pileup.php}}. Following these guidelines, we found
 that five pixels had to be excluded in the bright core to mitigate the effect of pile-up. The remaining six observations have $<0.5~\mathrm{ct~s^{-1}}$ and are not affected by pile-up (see Table~\ref{Log of}).

\section{Results}\label{sec:resu}

\subsection{X-ray spectral fitting}\label{subsec:x-ray}
To fit the X-ray spectra, we used {\textsc{xspec}}\footnote{\url{https://heasarc.gsfc.nasa.gov/xanadu/xspec/}} \citep[v.12.10.1;][]{1996ASPC..101...17A}.
Given the low count rates (Table~\ref{Log of}), we used simple power law (\textsc{pegpwrlw}) and black body  (\textsc{bbodyrad}) models to describe the data. For both models, we took into account interstellar extinction by including the \textsc{tbabs} model and used C-statistics due to low data counts. For this absorption model we used abundances set to those of \citet{Wilms2000} and the cross-sections from \citet{Verner1996}. Both models yielded the same quality of fit, so that we cannot statistically prefer one model over the other. However, in order to use the \citet{Wijnands2015} method to probe the nature of the compact accretor, we need to use the power-law model. Therefore, we here report on the results from fitting the absorbed power-law model, but we include the results for the absorbed black-body fits in  the Appendix for completeness.

%Although the source is very faint and there are not many counts in the XRT spectra, 
We also briefly explored whether the spectrum could be composed of two emission components, such as has been seen for VFXTs that have high-quality data available \citep[e.g.,][]{2011MNRAS.417..659A,2013MNRAS.434.1586A,2017MNRAS.464..398D}. For this we used the observation with the highest flux (observation ID 00031719004). We first fitted this to an absorbed powerlaw, then added a black body component and re-fitted. This resulted in similar fit parameters as for the single absorbed power-law model. The two-component model adequately fits the spectra by eye,
%(\textcolor{red}{$\chi^2$= 60} for 47 dof)
but the extra thermal component is not statistically required (F-test probability $>$ 0.99). It is likely that the low number of counts in the spectrum does not allow us to detect a second component, even if it is present. We therefore did not test this for the other observations, since these have even lower count rates. We conclude that a single-component model can adequately fit the \textit{Swift} spectra.

\begin{figure}
\centering
\includegraphics[width=1.0\textwidth]{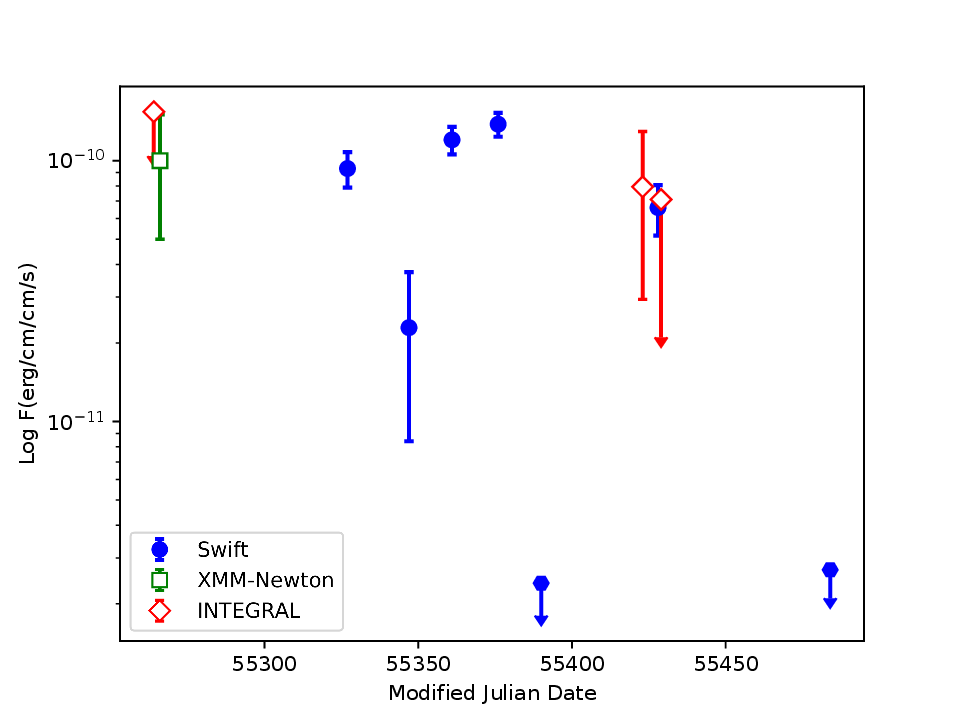}
\caption{\small Evolution of 0.5 -- 10 keV unabsorbed flux of \source, inferred from our spectral analysis of the {\it{Swift}}/XRT data (blue filled circles). To show the full outburst, we include points reported in the literature from \textit{INTEGRAL} (red open diamonds) and \textit{XMM-Newton} (green open square), which were converted to 0.5--10 keV for this purpose (see Section \ref{subsec:distance} and Table \ref{Log}).}
\label{1bear}
\end{figure}

\begin{figure}
\centering
%\rotatebox{270}{\includegraphics[width=0.66\textwidth]{fig/4peg24.png}} % Figure image
\includegraphics[width=1.0\textwidth]{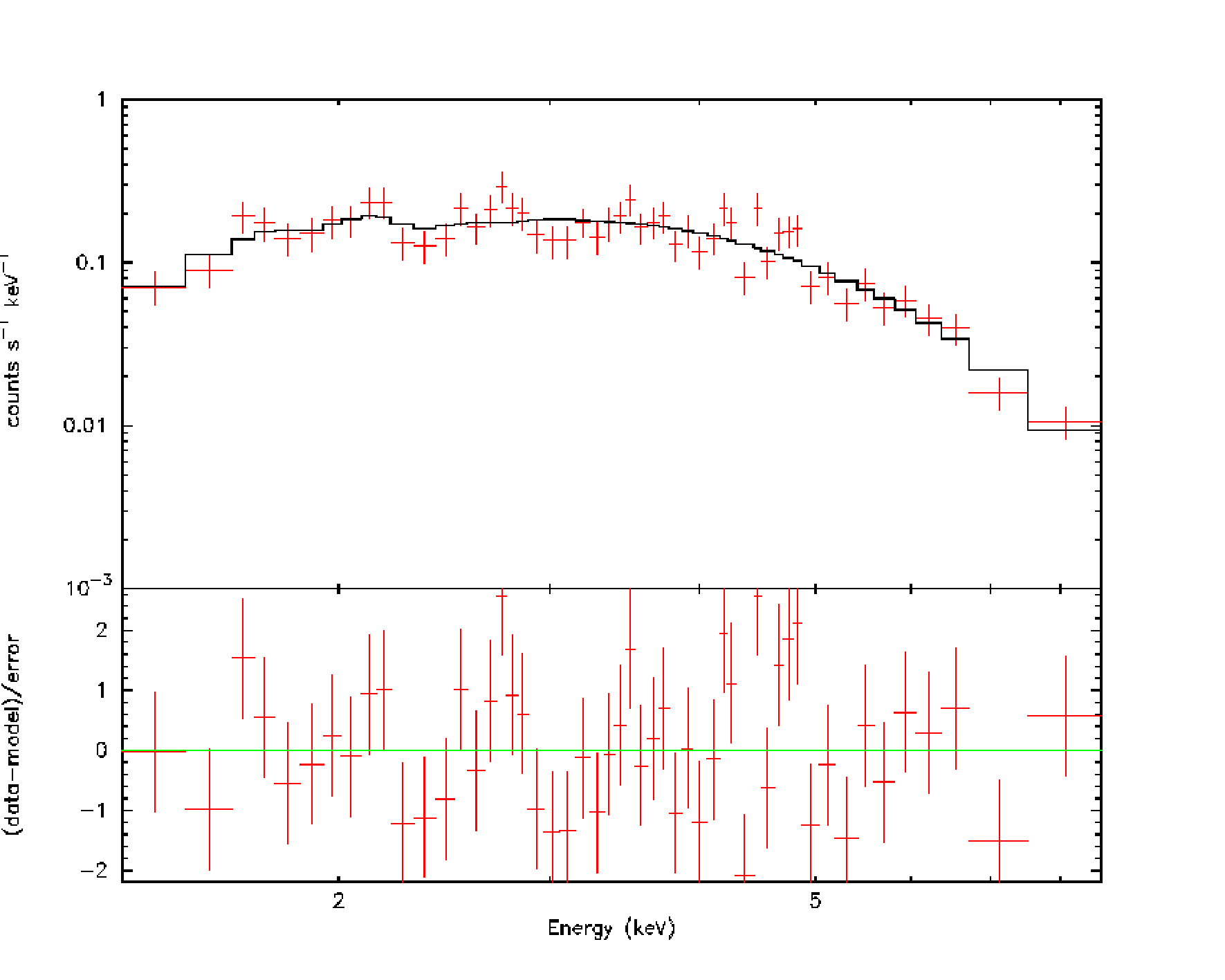} % 
\caption{\small An X-ray spectrum of \source\ detected with \textit{Swift}/XRT. Upper panel: shown is the brightest observation, 00031719004, fitted with an absorbed power law model. Bottom panel: the corresponding fit residuals in units of $\sigma$.}\label{21bear}
\end{figure}

% We calculated the absorbed flux from the power-law model with \textsc{cflux}\footnote{Using the syntax \textsc{cflux*tbabs*pegpwrlw} 
% Using the convolution model \textsc{cflux} in the energy range of 0.5 to 10 keV, fluxes reported are the unabsorbed fluxes within \textsc{xspec} and did the same for the unabsorbed flux. %\footnote{Using the syntax \textsc{tbabs*cflux*pegpwrlw} within \textsc{xspec}.}
Using the convolution model \textsc{cflux} within \textsc{XSpec}, setting  the energy range to 0.5 to 10 keV, we determined both the absorbed ($\mathrm{F}_{\mathrm{X,abs}}$) and unabsorbed fluxes ($\mathrm{F}_{\mathrm{X,unabs}}$). The results are listed in Table~\ref{tab:spec}. In Figure~\ref{1bear} we show the light curve constructed from the unabsorbed fluxes. From the seven XRT observations, the highest unabsorbed flux we measure is $\mathrm{F}_{\mathrm{X,unabs}}=13.8\times10^{-11}~{\mathrm{erg~cm}}^{-2}~{\mathrm{s^{-1}}}$ in observation 00031719004 (see Figure \ref{1bear} and Table \ref{tab:spec}). In Figure \ref{21bear} we show the \textit{Swift}/XRT spectrum of this observation.

\subsection{Flux upper limits for XRT non-detections}\label{subsec:nodetect}
In observations 00031719005 and 00031719007 the source was not detected with the XRT. For these observations, we determined the net counts detected at the source position with \textsc{xselect} (using similar source and background extraction regions as for the other observations; see Section \ref{subsec:descri}). For observation 00031719005 (1~ks) we detect 4 counts at the source position and 1 count averaged over the background regions. For observation 00031719007 (1.3~ks), we measure 6 counts for the source and none for the background. Accounting for small number statistics using the tables of \citet{gehrels1986}, we determine 95\% upper limit on the detected net source counts of 7.75 and 11.84 for observations 00031719005 and 00031719007, respectively. Dividing by the exposure times then gives 95\% confidence count rate upper limits of $<7.9\times 10^{-3}$~ct~s$^{-1}$ (00031719005) and $<9.1\times 10^{-3}$~ct~s$^{-1}$ (00031719007).%
%The image of observation 00031719005 has 256 counts for 0.26 ct/s with 1 ks exposure time and observation 00031719007 has 345 counts for 0.27 ct/s with 1.3 ks exposure time (eg., Table \ref{Log of}).
%This way we determined an upper limit on the net source count rate of $<0.006 ~\mathrm{ct/s}$ for observation 00031719005 and $<0.015 ~\mathrm{ct/s}$ for 00031719007. 
%

To estimate flux upper limits for the non-detections, we used \textsc{webpimms}\footnote{\url{https://heasarc.gsfc.nasa.gov/cgi-bin/Tools/w3pimms/w3pimms.pl}} to convert the count rate upper limits, assuming an absorbed power-law model with a photon index of $\Gamma = 2.49$ and a hydrogen column density of $\mathrm{N}_{\mathrm{H}}=5.81 \times 10^{22}~{\mathrm{cm}}^{-2}$. We choose those values because these are the ones we obtained for the observation with the lowest flux (observation 00031719002).

\subsection{X-ray spectral evolution}\label{subsec:distance}
\begin{figure}
\begin{center}
 \includegraphics[width=5.4cm, height=5cm]{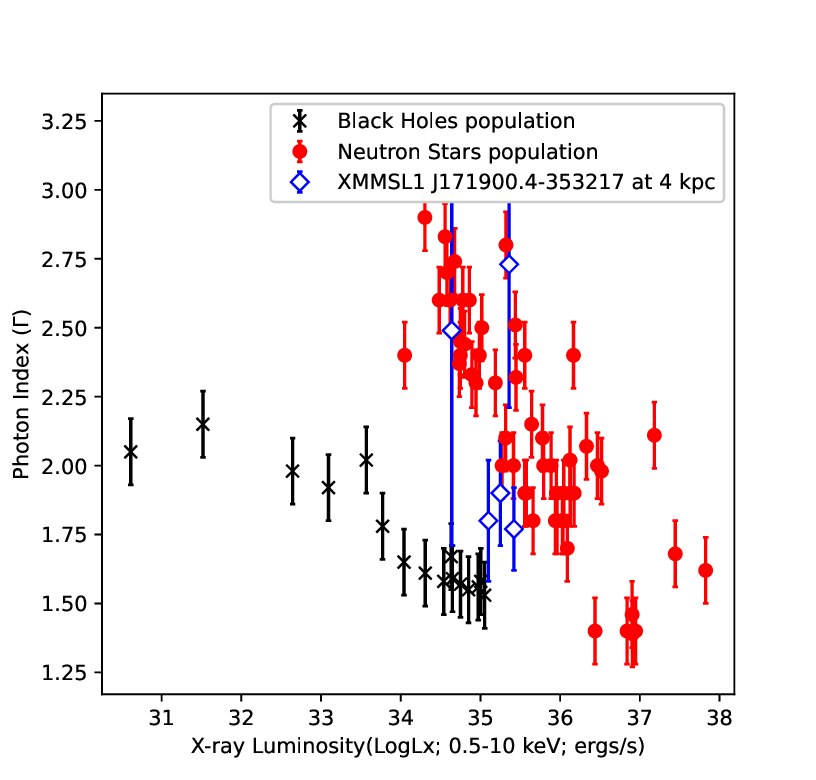}(a)\hfill
	\includegraphics[width=5.4cm, height=5cm]{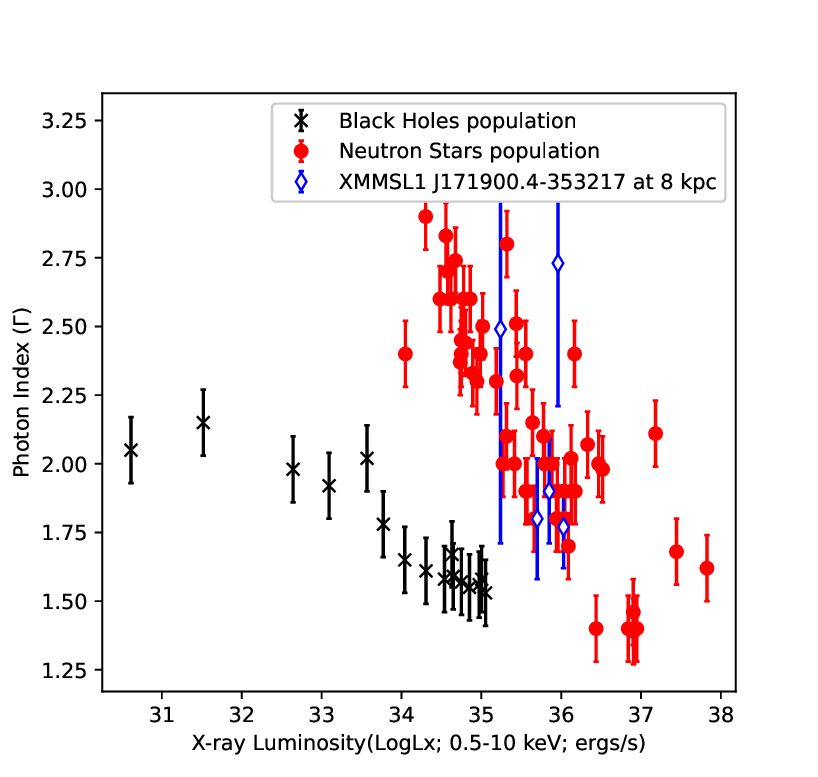}(b)\par 
	\includegraphics[width=7.8cm, height=6cm]{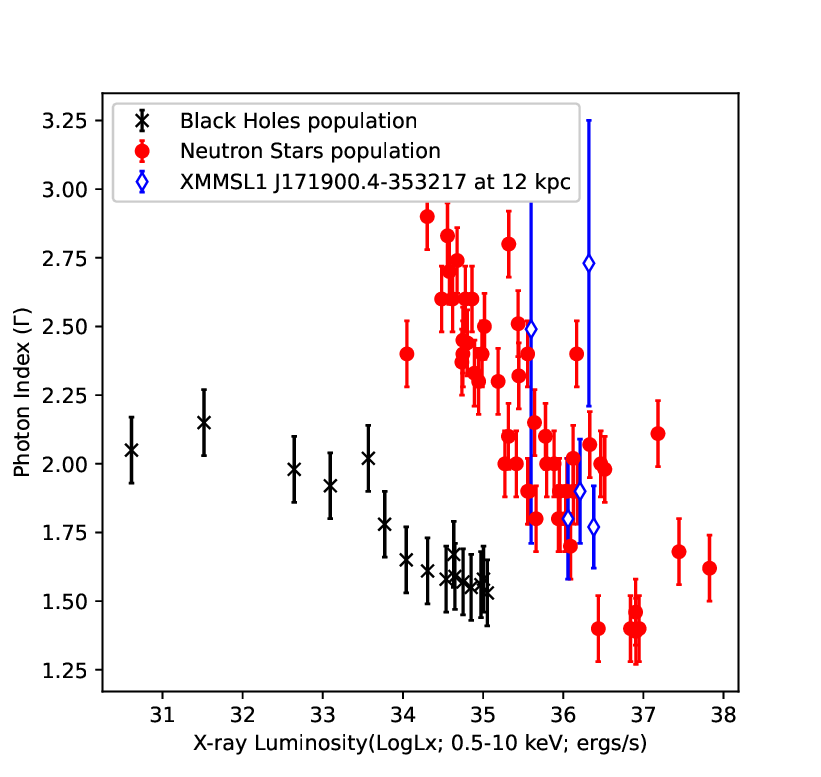}(c)
\end{center}	
\caption{\small Power-law index versus X-ray luminosity in the 0.5--10 keV range for \source\ as well as a sample of NS (red circles) and BH (grey crosses) LMXBs \citep[from][]{Wijnands2015}. For \source\ we used three different distances of 4 kpc (panel~a), 8 kpc (panel~b) and 12 kpc (panel~c). }
	\label{3bear}
\end{figure}

The distance of \source\ is unknown. We therefore took three different values of 4, 8 and 12 kpc, to calculate the 0.5--10 keV luminosity from the unabsorbed flux. These results are included in Table \ref{tab:spec}.
%The X-ray luminosity was calculated from the single absorbed power-law model with its errors using the relationship $L_{\mathrm{X}}=4\pi d^2F_{\mathrm{unabs}}$. 
In Figure \ref{3bear} we plot the evolution of the power-law index versus luminosity of \source\ along with the sample of NS (red filled circles) and BH (black crosses) LMXBs of \citet{Wijnands2015}. 
%The circular data points in Figure \ref{3bear} are NS systems and the crosses are BH transients \citep[taken from][]{Wijnands2015}. 
We then overplot \source\ as blue open diamonds for different distances of 4, 8, and 12 kpc in sub-panels a, b, and c, respectively. 
%In sub-panel (a) using 4 kpc we see the source evolve over luminosities in the range $\simeq5\times10^{34}--10^{35} ~{\mathrm{erg/ s}}$ (thin-diamond points) in panel (b) for 8~kpc we have luminosities in the range $\simeq10^{35}--10^{36} ~{\mathrm{erg/s}}$ (thin-diamond points), and finally in panel (c) luminosities in the range $\simeq5\times10^{35}--10^{36} ~{\mathrm{erg / s}}$ for a distance of 12 kpc.   

We find that for all distances, our data points fall among the NS sample, but above the BH track. This would suggest that the source is either a proximate ($\lesssim$ 4 kpc) BH, or a NS located around or beyond 4 kpc. Considering the high $\mathrm{N_H}$ towards the source, both as inferred from our X-ray spectral fitting ($\mathrm{N}_{\mathrm{H}}\simeq 5\times 10^{22}~{\mathrm{cm}}^{-2}$) and from Galactic extinction maps ($\mathrm{N}_{\mathrm{H}}\simeq1\times10^{22}~{\mathrm{cm}}^{-2}$;\citet{bekhti2016hi4pi}), we consider a larger distance more likely, and hence tentatively favor a NS nature. However, the reader should bear in mind that a BH nature cannot be excluded.

\subsection{Time-averaged accretion rate}
We continue to calculate the time-averaged accretion rate for \source, since this is an interesting parameter to understand the possible evolution paths of VFXTs \citep[][]{King2006}. We initially assume that the source contains a NS primary and then calculate the mean outburst accretion rate, $\langle \dot{\mathrm{M}}_{\mathrm{ob}}\rangle$, from the mean unabsorbed flux measured during the outburst. For this purpose, we add the \textit{INTEGRAL} and \textit{XMM-Newton} fluxes reported in the literature to our results obtained with \textit{Swift}. 

We used \textsc{webpimms} to convert reported instrument count rates or 20--40 keV fluxes to unabsorbed 0.5--10 keV fluxes. All information used for these conversions are listed in Table~\ref{Log}. We assumed an absorbed power-law spectral shape. For the \textit{XMM-Newton} and first \textit{INTEGRAL} observations, both performed in March 2010, we used $\mathrm{N}_{\mathrm{H}}=5.45\times 10^{22}~{\mathrm{cm}}^{-2}$ and $\Gamma=1.90$, which are the values we obtained from spectral fitting for the \swift/XRT observations performed closest in time (observation 00000031719001; see Tables~\ref{Log of}--\ref{Log}).
%We used th. For the \textit{INTEGRAL} data we used 20--40 keV input and 0.5--10 keV output energy ranges. As shown in Table \ref{some} the \textit{INTEGRAL} count rate data from \citet{2010ATel.2616....1B} was converted using spectral parameters of neutral hydrogen column density $N_{\mathrm{H}}=5.04\times 10^{22}/\mathrm{cm}^2$ and $\Gamma=1.78$ because we estimate of the closest outburst with \textit{Swift} observation 00000031719001 once again. %We used an online tool \footnote{\url{https://www.dsf.unica.it/~riggio/calcs.html}} to convert 8 mCrab to the count rate from 20 keV minimum energy to 40 keV maximum energy ranges.
For the other two \textit{INTEGRAL} observations, both performed in 2010 August, we assumed $\mathrm{N}_{\mathrm{H}}=6.67\times 10^{22}~{\mathrm{cm}}^{-2}$ and $\Gamma=1.80$ as found from fitting the \textit{Swift}/XRT spectrum obtained closest in time (observation 00000031719006).

The resulting 0.5--10 keV flux light curve 
%obtained in this way, running from the first detection with %\textit{XMM-Newton} on 2010 March 10 till the last %\textit{Swift} observation in 2010 October 15,
is shown in Figure \ref{1bear}. From all these data points we determine a mean 0.5--10 keV  outburst flux of $8.9\times10^{-11}~{\mathrm{erg~cm}}^{-2}~{\mathrm{s^{-1}}}$. Based on this, we estimate the 0.1--100 keV accretion luminosity by assuming a bolometric correction factor of 3  \citep[following][]{2007A&A...465..953I}. 
The mass transfer rate of the outburst was then computed using the equation $\langle \dot{\mathrm{M}}_{\mathrm{ob}}\rangle=\mathrm{R}_{\mathrm{Ns}}\mathrm{L}_{\mathrm{acc}}/\mathrm{GM}_{\mathrm{Ns}}$, where $\mathrm{G}=6.67\times 10^{-8}~{\mathrm{cm}}^3{\mathrm{g}^{-1}}{\mathrm{s}}^{-2}$ is the gravitational constant.
Assuming $\mathrm{R}_{\mathrm{Ns}}=1.1\times10^6 ~{\mathrm{cm}}=11~{\mathrm{km}}$ and $\mathrm{M}_{\mathrm{Ns}}=1.5~\mathrm{M}_{\odot}$, the outburst mass accretion rate we obtain is $\langle \dot{\mathrm{M}}_{\mathrm{ob}}\rangle \simeq 1.7 \times10^{-10}~\mathrm{M}_{\odot}~{\mathrm{yr}^{-1}}$.

We can next estimate the mean long-term averaged accretion rate using $\langle\dot{\mathrm{M}}_{\mathrm{long}}\rangle=
  \langle \dot{\mathrm{M}}_{\mathrm{ob}}\rangle \times \mathrm{t}_{\mathrm{ob}}/\mathrm{t}_{\mathrm{rec}}$, where $\mathrm{t}_{\mathrm{ob}}$ is the outburst
duration, $\mathrm{t}_{\mathrm{rec}}$ is the system's recurrence time, and the ratio of the two represents its duty cycle. Neither the onset nor the fading of the outburst into quiescence have been observed for \source, so the total outburst duration is unconstrained.\footnote{We note that in the \rxte/PCA bulge scans the possibly associated transient XTE J1719--356 seems to be detected on and off between 2010 March and September, but not thereafter. The \rxte\ data therefore does not provide additional constraints on the outburst duration. See https://asd.gsfc.nasa.gov/Craig.Markwardt/galscan/html/ XTE$\_$J1719-356.html} If we assume that the source was continuously active (i.e., only occasionally dropping to non-detectable flux levels) between its first and last detection on 2010 March 9 and 2010 August 20, the minimum outburst duration is 164~days. Since this was the first and only outburst ever observed for the source, its recurrence time is also unconstrained. For the present purpose we assume a duty cycle of 1--10\% based on long-term X-ray monitoring of VFXTs in the Galactic center \citep[][]{degenaar2009behavior,degenaar2010four}. This would imply an outburst recurrence time of 4.5--45~yr for \source, and yields a mean long-term accretion rate of $\langle\dot{\mathrm{M}}_{\mathrm{long}}\rangle \simeq 0.17-1.7 \times10^{-11}~\mathrm{M}_{\odot}~{\mathrm{yr}^{-1}}$.  

We note that if the source harbors a black hole accretor, the above estimates for the (long-term) mass accretion rate would be a factor $\simeq10$ lower due to the mass difference between neutron stars and black holes.

\section{Discussion}\label{sec:discuss}
We report on the properties of the discovery outburst of the X-ray transient \source, which lasted more than 164 days in 2010. We studied the X-ray spectral evolution of the source using the Swift/XRT data and used the method of \citet{Wijnands2015} to investigate the nature of the accreting object. Based on the evolution of its power-law index with 0.5--10 keV luminosity, we conclude that \source\ is most likely a NS LMXB located at several kpc.

%  We have tested our result with the original BH and NS data sample in \citet{Wijnands2015} and the result most resemble with the NS. Even though, our calculated luminosity is slightly position dependent the the candidate \source\ is most-likely a NS X-ray transient at nearly the center of our galaxy and at 12 kpc (see Figure \ref{3bear}). At those positions the luminosity of X-ray transient \source\ become soften at $10^{36} ~{\mathrm {erg/s}}$ which coincides with the NS X-ray transients of model sample data\citep[][]{Wijnands2015}.

Adding to our \swift/XRT results flux measurements reported in the literature (from \xmm\ and \integral\ observations), we constructed the light curve of the 2010 outburst (see Figure \ref{Log of}). Over the 5.5~months that the source was observed to be active, the maximum 0.5--10 keV unabsorbed flux detected with \textit{Swift}/XRT was $\mathrm{F}_{\mathrm{X,unabs}}^{\mathrm{peak}}=13.8\times 10^{-11} ~{\mathrm{erg~cm}}^{-2}~{\mathrm{s^{-1}}}$. For a distance of 8 kpc, this peak flux translates into a luminosity of $\mathrm{L}_{\mathrm{X}}^{\mathrm{peak}} \simeq 1.1\times 10^{36} ~{\mathrm{erg~s^{-1}}}$. This classifies \source\ as a VFXT.\footnote{We note that \citet{Wijnands2006} uses the 2--10 keV band to define luminosity classes whereas we here use 0.5--10 keV. However, since the 0.5--10 keV luminosity is higher than the 2--10 keV luminosity, our conclusion that \source\ falls in the regime of VFXTs still holds.} % (see Figure \ref{1bear}).

We furthermore estimated a mean unabsorbed flux along the observations of $\mathrm{F}_{\mathrm{X}}^{\mathrm{avg}} \simeq 8.9\times10^{-11}~ {\mathrm{erg~cm}}^{-2}~{\mathrm{s^{-1}}}$. For a distance of 8~kpc, this translates into a luminosity of $\mathrm{L}_{\mathrm{X}}^{\mathrm{avg}} \simeq 6.8\times10^{35}~{\mathrm{erg~s^{-1}}}$. We used this information to estimate the average accretion rate along the outburst as $<\dot{\mathrm{M}}_{\mathrm{ob}}> \simeq 1.7 \times10^{-10}~\mathrm{M}_{\odot}~{\mathrm{yr}^{-1}}$. If the source has a duty cycle of 1--10\%, which is not uncommon for LMXBs and VFXTs \citep[][]{degenaar2010four}, we can then estimate a long-term average accretion rate of $<\dot{\mathrm{M}}_{\mathrm{long}}> \simeq 0.17-1.7 \times10^{-11}~\mathrm{M}_{\odot}~{\mathrm{yr}^{-1}}$.  This is in the same range as inferred for the VFXTs in the Galactic Center \citep[][]{degenaar2009behavior,degenaar2010four}.
% luminosity of $5.5\times 10^{35} ~{\mathrm{erg/s}}$. }
Very low long-term average accretion rates can only be explained if these systems have hydrogen poor companions or are born with low companion masses \citep[][]{King2006}. However, the current (faint) accretion activity may not necessarily be representative for the long-term behavior of these systems \citep[][]{wijnands2013}.

In the past years, several NS LMXBs with similarly low outburst luminosities (hence accretion rates) as \source\ were uncovered to harbor accreting millisecond X-ray pulsars (AMXPs). Examples are IGR J17062--6143 \citep[][]{strohmayer2017}, IGR J17591--2342 \citep[][]{sanna2018}, IGR J17379--3747 \citep[][]{sanna2018-2} and
IGR J17494--3030 \citep[][]{ng2020}. All were previously known VXFTs that were observed during (new) outbursts with \nicer, which detected the X-ray pulsations. Given the similar X-ray spectral properties of \source\ with those sources, we hypothesize that it may also harbor a millisecond X-ray pulsar. Indeed, one of the sources mentioned above, was proposed to be a NS based on the same method as we employ here \citep[][]{armaspadilla2013} and later found to be an AMXP \citep[][]{ng2020}. Therefore, should \source\ enter a new accretion outburst in the future, we encourage X-ray observations (in particular with \nicer) to search for pulsations that would confirm the NS nature of this source and allow for a measurement of its orbital period. In case a new outburst occurs, we also encourage dense monitoring of the outburst decay (in particular with \swift), since this can also provide an indication of the orbital period and nature of the compact accretor \citep[e.g.,][]{2011MNRAS.417..659A,heinke2015,stoop2021}.

\section*{ACKNOWLEGEMENTS}
OA is grateful to Sera Markoff and the Anton Pannekoek Institute for organizing and hosting the Advanced Theoretical Astrophysics summer school in 2019, which fostered the collaboration that led to this work. ND was partly supported by a Vidi grant awarded by the Netherlands organization for scientific research (NWO). This work made use of data supplied by the UK \swift\ Science Data Centre at the University of Leicester. M. A. P. acknowledges support from the Spanish ministry of science under grant PID2020--120323GB-I00. M. A. P. acknowledges support from the Consejeria de Economia, Conocimiento y Empleo del Gobierno de Canarias and the European Regional Development Fund under grant ProID2021010132.

\begin{appendices}
\section{Black-body spectral fitting results}
\label{sec:A}
For completeness we here report on the results of fitting the \swift/XRT spectra of \source\ with an absorbed black body model. For the upper limit calculation of the two XRT non-detections, we now used $\mathrm{kT}=1.06$~keV and $\mathrm{N}_{\mathrm{H}}=2.29 \times 10^{22} ~{\mathrm{cm}}^{-2}$. These are the values we obtained for the observation with the lowest flux (observation ID 00031719002). All results are listed in Table \ref{Flux}.

\begin{table*}[!t]\centering
  \small
  \newcommand{\DS}{\hspace{6\tabcolsep}} %% Expanded Space between
  %% some cols
  \begin{changemargin}{-2cm}{-2cm}
    \caption{Results from XRT spectral analysis using a black-body model. X-ray fluxes and luminosities are given in the 0.5--10 keV energy band.} \label{Flux}
    \setlength{\tabnotewidth}{0.95\linewidth}
    \setlength{\tabcolsep}{0.7\tabcolsep} \tablecols{10}
    \begin{tabular}{l @{\DS} cccc l cccc}
      \toprule
Obs & N$_{\mathrm{H}}$& kT&    F$_{\mathrm{X}_\mathrm{abs}}$ & F$_{\mathrm{X}_\mathrm{unabs}}$ & L$_{\mathrm{X}}$ 4 kpc & L$_{\mathrm{X}}$ 8 kpc & L$_{\mathrm{X}}$ 12 kpc\\
& ($10^{22}~{\mathrm{cm}}^{-2}$)& (keV) & \multicolumn{2}{c}{($10^{-11}~ {\mathrm{erg~cm}}^{-2}{\mathrm{s}^{-1}}$}) &  \multicolumn{3}{c}{($10^{35} ~{\mathrm{erg~s^{-1}}}$)}\\
\midrule
1 & $2.23\pm0.40$ & $1.32^{+0.08}_{-0.07}$ & $3.71^{+0.18}_{-0.16}$ & $4.47\pm0.20$ & $0.85^{+0.04}_{-0.03}$&$3.42^{+0.16}_{-0.18}$ & $7.70^{+0.36}_{-0.35}$\\
2 & $2.29^{+1.56}_{-1.29}$ & $1.06^{+0.24}_{-0.19}$ & $0.50^{+0.10}_{-0.08}$ & $0.65^{+0.12}_{-0.11}$ & $0.12^{+0.03}_{-0.02}$&$0.51\pm0.09$  & $1.12^{+0.21}_{-0.19}$\\
3 & $3.38^{+1.26}_{-1.19}$ & $1.01^{+0.13}_{-0.11}$ & $1.74^{+0.21}_{-0.16}$ & $2.51^{+6.61}_{-1.72}$ & $0.48^{+1.26}_{-0.33}$ & $1.92^{+5.06}_{-1.32}$ & $4.32^{+11.39}_{-2.96}$ \\
4 & $1.69^{+0.31}_{-0.30}$ & $1.34^{+0.07}_{-0.06}$ & $6.17^{+0.29}_{-0.28}$ & $7.24^{+0.35}_{-0.32}$ & $1.38^{+0.07}_{-0.06}$ & $5.54^{+0.27}_{-0.24}$ & $12.47^{+0.6}_{-0.55}$ \\
5 & 2.29 fix & 1.06 fix &$<0.05$ & $<0.07$ &$<0.02$ & $<0.06$ & $<0.13$ \\
6 & $2.96^{+0.60}_{-0.57}$ & $1.41^{+0.10}_{-0.09}$  & $2.63\pm0.18$ & $3.24\pm0.15$ & $0.62\pm0.03$&$2.48^{+0.12}_{-0.11}$ & $5.58\pm0.26$\\
7 & 2.29 fix & 1.06 fix & $<0.06$ & $<0.08$ &$<0.02$ & $<0.07$ & $<0.15$\\
\bottomrule
\tabnotetext{}{Quoted errors reflect $1-\sigma$ confidence intervals.}
\end{tabular}

%\tablenotetext{a}{Neutral hydrogen column density in units of $10^{22}/{\mathrm{cm}}^{2}$}
% \tablenotetext{b}{Flux (in units of $10^{-12}~ {\mathrm{erg/cm}}^{2}/{\mathrm{s}}$ over the X-ray energy 0.5 -- 10 keV)}
% \tablenotetext{c}{nhp(null hypothesis probability)-probability that the model is correct for those data points (if close to unity)}
%\tablenotetext{b}{Flux in units of $10^{-11}~ {\mathrm{erg/cm}}^{2}/{\mathrm{s}}$}
%\tablenotetext{c}{X-ray luminosity in units of $10^{35} ~{\mathrm{erg/s}}$ calculated from the unabsorbed flux by adopting distances of 4 kpc, 8 kpc and 12 kpc.}
\end{changemargin}
\end{table*}
\end{appendices}
\bibliographystyle{rmaa}
\bibliography{refs}
\end{document}